\documentclass[aps,pra,twocolumn,showpacs]{revtex4-1}

\usepackage{amsfonts,amssymb,amsmath}
\usepackage{mathtools}
\usepackage[]{graphics,graphicx,epsfig}
\usepackage{amsthm,multirow}
\usepackage{color}
\begin{document}

\title{Quantum-correlation-based free-space optical link with an active reflector}

\author{Dongkyu Kim}
\affiliation{Emerging Science and Technology Directorate, Agency for Defense Development, Daejeon 34186, South Korea}

\author{Dohoon Lim}
\affiliation{Emerging Science and Technology Directorate, Agency for Defense Development, Daejeon 34186, South Korea}

\author{Kyungdeuk Park}
\affiliation{Emerging Science and Technology Directorate, Agency for Defense Development, Daejeon 34186, South Korea}

\author{Yong Sup Ihn}
\email{yong0862@add.re.kr}
\affiliation{Emerging Science and Technology Directorate, Agency for Defense Development, Daejeon 34186, South Korea}

\date{\today}

\begin{abstract}
We present a quantum-correlation-based free-space optical(FSO) link over 250 m using an outdoor active reflector 125 m from the transceiver station. 
The performance of free-space optical communication can be significantly degraded by atmospheric turbulence effects, such as beam wander and signal fluctuations. 
We used a 660 nm tracking laser to reduce atmospheric effects, by analyzing the fast beam wander and slow temporal beam drift, using this information to correct the quantum channel alignment of the 810 nm signal photons. 
In this work, the active reflector consisted of a mirror, a 6-axis hexapod stage, and a long-range wireless bridge. 
The slow drift of the beam path due to outdoor temperature changes was steered and controlled using wireless optical feedback between the receiver units and the active reflector. 
Our work provides useful knowledge for improved control of beam paths in outdoor conditions, which can be developed to ensure high quality quantum information transfer in real-world scenarios, such as an unmanned FSO link for urban quantum communication or retro-reflective quantum communication links.
\end{abstract}

\maketitle

\section{Introduction}
Quantum key distribution (QKD) offers a method of distributing and sharing encryption keys among known users. 
Thus far, enormous advances have been made in QKD from ground-to-ground, plane-to-ground, and ground-to-satellite transmission \cite{Nature07Ursin,QST17Pugh,Science17Yin}. 
As a result, FSO systems have naturally attracted considerable attention as a critical technology for large-scale quantum networks, QKD in urban areas, retro-reflective quantum communication links, and unmanned mobile FSO links \cite{AO13Garcia,SPIE06Kullander,OE18Ravinovich,SPIE19Quintana,IEEE21Alshaer,SPIE22Conrad}. 
Moreover, optical fibers and free-space are the most used transmission channels for both classical and quantum communication systems. 
While optical-fiber-based links are usually buried and can be affected by data capacity and inaccessibility of points, FSO links are wirelessly installed and portable. 
They can also provide higher bandwidths than radio-frequency (RF)-based communications.
In this respect, FSO links are useful in a metropolitan networks affected by mass data traffic, and also  free-space QKD would be of interest to various military scenarios to distribute secure cryptographic keys. 

Nevertheless, because it must operate outdoors, several technical issues need to be considered. 
Because of the strong background noise from scattered sunlight, a free-space optical link should be designed with a highly stable coupling of the signal photons into single-mode fibers (SMFs) under atmospheric turbulence \cite{SR18Ko,npjQI21Avesani}. 
Moreover, atmospheric turbulence can lead to various detrimental effects$-$including optical loss, beam wander, arrival-angular fluctuation, and received-signal fluctuations, which result in a reduction in the signal-to-noise ratio (SNR) and an increase in the the quantum bit error rate (QBER) in the quantum communication process \cite{SPIE02Davis,OptEn14Casado,IEEE18Fernandez}.
Consequently, proper optical feedback is required to compensate for such turbulence effects and to maximize the signal photon-fiber coupling in the quantum channel.
So far, various kinds of methods to correct atmospheric effects have been proposed.
By using a fast steering mirror (FSM) which can work at high frequency under the atmospheric turbulence, the jitter error caused by atmospheric turbulence can be compensated tracking the spot position \cite{OptEn14Casado,IEEE18Fernandez,OE12Takenaka,IEEEPH18Li}.
In another method to use a deformable mirror, the coupling efficiency was improved correcting the wavefront distortion caused by atmospheric turbulence \cite{AO15Chen}.
In addition, various correction methods using dual-wedges and prisms have been introduced \cite{OE16Li,OFT21Cao}.

In this work, we have experimentally demonstrated a 250 m free-space optical link based on a non-classical photon-pair source.
In the transmitter stage, the quantum signal is combined with a weak tracking laser, both beams being sent to an active reflector 125 m away from the transceiver station. 
The refelcted signal photons are then correctly aligned to a SMF in the quantum channel receiver using a closed-loop tracking system and a wirelessly-controlled hexapod stage.  

The paper is organized as follows: Section II presents the overall experimental setup, including a 250 m free-space optical link with an active reflector, spontaneous parametric down-conversion (SPDC) quantum source, and monitoring setup for atmospheric effects; Section III discusses the performance of the optical feedback setup consisting of a FSM and position-sensitive detectors (PSDs) in a beam tracking system and the experimental results of the 250 m free-space optical link, which is summarized in Section IV. 

\begin{figure*}
\includegraphics[width=0.75\textwidth]{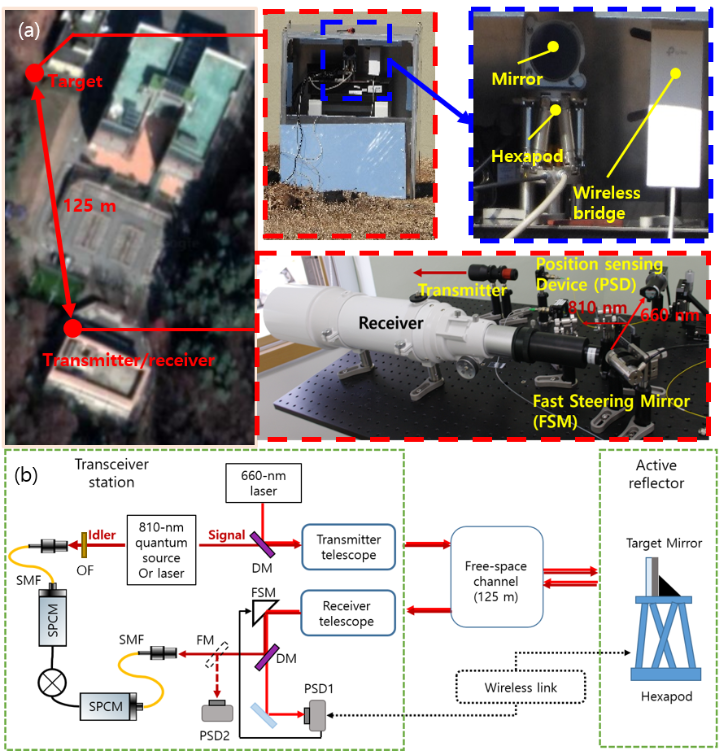}
\caption{(a) Image and (b) schematic diagram of the 250 m FSO link setup with an active reflector. The transmitter and receiver units are installed on the same optical board. Quantum signal photons at 810 nm and a 660 nm weak tracking laser are retro-reflected form an active reflector 125 m from the transmitter and receiver station. SMF, single-mode fiber; OF, optical filter; SPCM, single-photon counting module; DM, dichroic mirror; FM, flippery mirror; FSM, fast steering mirror; PSD, position-sensitive detector.
}
\label{fig:fig1}
\end{figure*}

\section{Experimental setup}
The overall experimental setup for a 250 m free-space optical link with an active reflector is shown in Fig. \ref{fig:fig1}.
The transceiver station and active reflector are linked by a free-space channel established between the quantum research building and outdoor aluminum box (Fig. \ref{fig:fig1}(a)).
The transmitter consists of a fiber collimator and an achromatic Galilean beam expander (Thorlabs, GBE10-B) with 10$\times$ fixed magnification.
The receiver was a Galilean telescope (Sky Explorer, SE120) with a focal length of 600 mm comprising an objective lens of a diameter 120 mm.
The transmitter and receiver units were installed on the same optical breadboard and mounted on a 6-axis hexapod (PI, H-850-H2).
A collimated 15 mm diameter beam was transmitted to the receiver via an active reflector seperated by 125 m.
To characterize and correct the beam wander of the received beam, the system uses a laser of 660 nm wavelength, referred to as the tracking laser, and the 810 nm wavelength for carrying single photons, referred to as the quantum channel.
The 810 nm photon-pairs allow us to use Si-based single photon detectors (SPDs), which have relatively high detection efficiency, and generate the bright photon-pairs using the type-0 SPDC process in a ppKTP crystal.

Figure \ref{fig:fig1}(b) shows a schematic of the proposed system.
The signal photons of the quantum channel, generated from a SPDC source, are combined with a weak 660 nm tracking channel laser using a dichroic mirror.
Both beams are transmitted through a 125 m free-space channel and reflected using a target mirror mounted on a small 6-axis hexapod.
The returned signal photons and tracking laser are directed to a closed-loop optical feedback system through the receiver telescope.
Turbulence effects can be monitored in real-time using PSDs for the closed-loop feedback configuration.
This is because the FSM is placed before the quantum and tracking channels, which should be located at the same distance.
The FSM is PID (proportional-integral-derivative)-controlled using the centroid position of a weak 660 nm tracking laser obtained by PSD1.
Along with controlling the PID of the FSM, PDS1 and the active reflector hexapod remotely control the slow beam drift owing to the outdoor temperature. 
A flipper mirror (FM) and PSD2 can them be used to monitor the centroid position of a weak 810-nm laser with the same wavelength as the SPDC source.
Signal photons in the quantum channel are then coupled to an SMF and detected using a single photon counting module (SPCM, Excelitas).
Finally, the cross-correlation function, $g^{(2)}(\tau)$, between the idler and received signal photons is measured.
The background noise was blocked by an optical filter (Thorlabs, FB800-40) with a center wavelength of 800 nm and a full-width at-half-maximum (FWHM) of 40 nm, which is sufficient to accept signal photons with a center wavelength of 810 nm.
\begin{figure}
\includegraphics[width=0.48\textwidth]{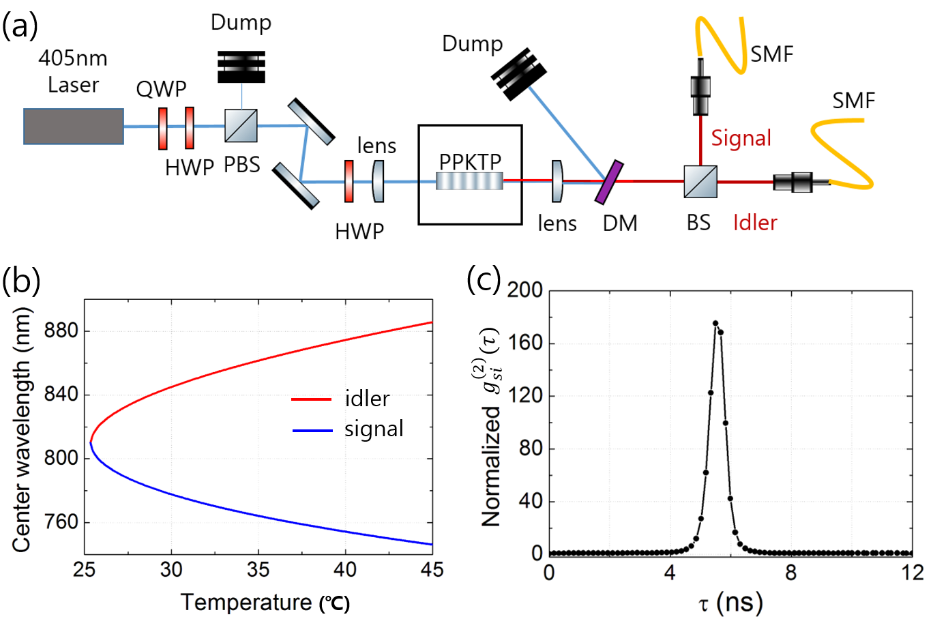}
\caption{(a) Experimental setup for the broadband laser diode (LD) pumped type-0 ppKTP SPDC source. (b) Calculated phase-matching curve as a function of the temperature. (c) Cross-correlation function between the degenerated signal and idler photons generated from a 30 mm-long ppKTP. QWP, quarter-wave plate; HWP, half-wave plate; PBS, polarizing beam splitter; BS, beam splitter.
}
\label{fig:fig2}
\end{figure}

\begin{figure}
\includegraphics[width=0.44\textwidth]{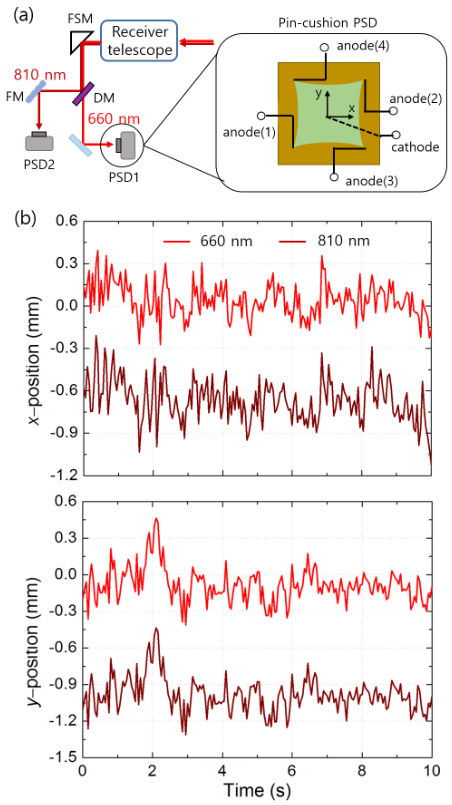}
\caption{(a) Experimental setup for measuring beam wander depending on the wavelength. The inset shows a pin-cushion type PSD. (b) x and y$-$positions of the 660 nm and 810 nm beams undergoing beam wander. Individual data sets are vertically offset for clarity.}
\label{fig:fig3}
\end{figure}

\subsection{Photon-pair source}
Broadband multi-mode laser diode (LD) at 405 nm ($\Delta\lambda\sim0.5$ nm) was used as the pump laser.
Degenerate photon-pairs with a center wavelength of 810 nm are collinearly generated via a type-0 ($e \to e+e$) SPDC process using a 30 mm-long periodically-poled potassium titanyl phosphate (ppKTP, Racol) crystal with a 3.425 $\mu$m poling period (Fig. 2(a)).
The SPDC process generally has the highest conversion efficiency when the pump, signal, and idler modes satisfy energy conservation and phase-matching conditions:
\begin{equation}
\begin{split}
\Delta\omega&=\omega_{p}-\omega_{s}-\omega_{i},\\
\Delta k&=k_{p}-k_{s}-k_{i}-\frac{2\pi}{\Lambda},
\label{eq1}
\end{split}
\end{equation}
where $\omega_{j=p,s,i}$ and $k_{j=p,s,i}$ are the angular frequency and wavenumber of the pump, signal, and idler modes, respectively.
Figure 2(b) shows the temperature dependence of the phase-matching curve calculated using Eq. \ref{eq1} \cite{OE12Steinlechner,JOSAB14Steinlechner}.
The phase-matching curve shows an increased non-degeneracy of the signal and idler photons as the temperature increases.
The pump beam was collimated to a diameter of 700 $\mu$m at FWHM.
A plano-convex lens of focal length, $f=300$ mm is then used to focus the pump beam to have the waist $\it{w}_{o}$ at $\sim$ 44 $\mu$m at the center of the crystal \cite{PRA05Ljunggren,PRA10Bennink}.
At the set temperature of 25.3 $^{\circ}$C, degenerate photon-pairs are produced with a waist size of $\sim$ 33 $\mu$m and propagate collinearly with the pump beam, blocked by a dichroic mirror (DM).
The signal and idler photons are separated using the BS and coupled to the SMFs.
The idler photons are directly sent to an SPCM to herald the signal photons sent to the target mirror through the telescope. 
The number of photon-pairs is aproximatedly 3.0$\times$10$^5$ Hz/mW in the coincidence counts.
Figure \ref{fig:fig2}(c) shows the normalized cross-correlation function $g^{(2)}(\tau)$ as a function of the delay time $\tau$.

\begin{figure}
\includegraphics[width=0.44\textwidth]{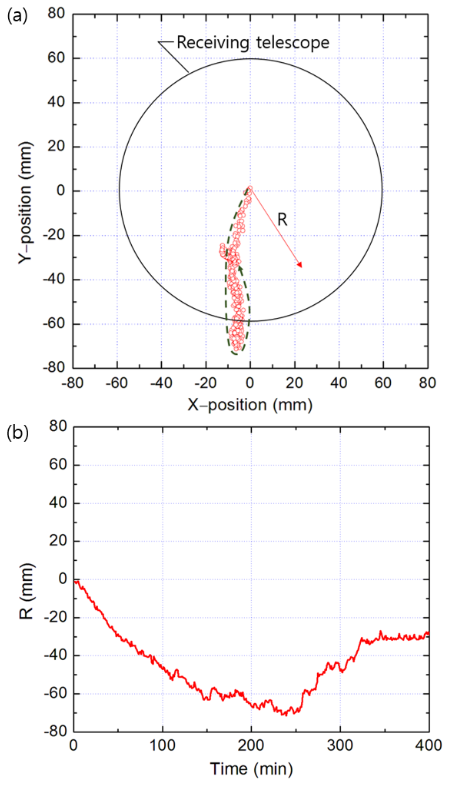}
\caption{The temporal drifts of (a) the beam centroid position and (b) the radial distance, $\left\vert\text{R}=\sqrt{\text{X}{^2}+\text{Y}{^2}}\right\vert$, over a period of 400 min. Each beam centroid position is measured every min. 
}
\label{fig:fig4}
\end{figure}

\begin{figure*}
\includegraphics[width=0.75\textwidth]{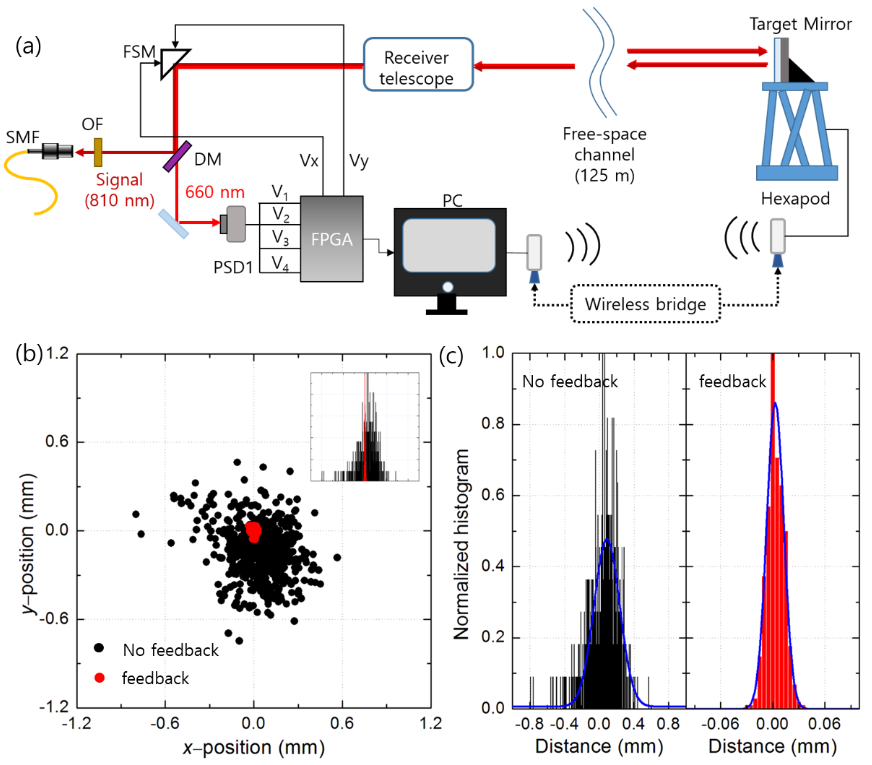}
\caption{(a) Schematic of beam wander correction for the 250 m free-space optical link in a closed-loop feedback configuration. (b) Beam centoid position of the 660 nm tracking laser measured using the PSD1 without and with beam correction. The inset shows a normalized histogram of the beam centroid radial positions. (c) Gaussian-fit curves (blue line) for normalized histograms are shown in the inset of Fig. \ref{fig:fig5}(b). All data are measured over a period of 30 s.}
\label{fig:fig5}
\end{figure*}

\begin{figure}
\includegraphics[width=0.44\textwidth]{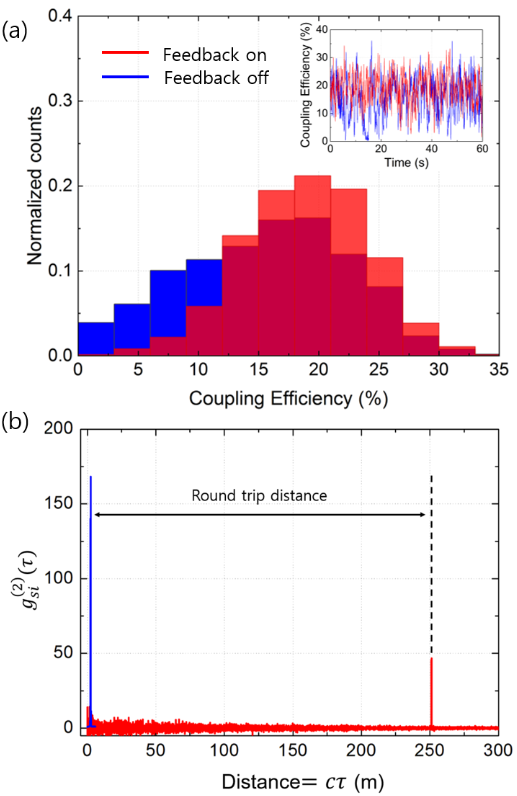}
\caption{(a) Histogram showing the statistical distribution of the SMF coupling efficiency in the quantum channel. A weak 810 nm laser is used for measurement. The inset shows the efficiency of the power coupled into the SMF. (b) Cross-correlation function between idler and signal photons in a 250 m free-space channel. $c$ is the speed of light.
}
\label{fig:fig6}
\end{figure}

\subsection{Atmospheric effects}
FSO links involve absorption and scattering depending on atmospheric conditions and transmission.
Therefore, one of the major challenges facing FSO systems is compensating for the effects of turbulence-induced photon fluctuations on system performance.
Atmospheric turbulence results from both the spatial and temporal fluctuations of the refractive index due to temperature, pressure, and wind variations along the optical path \cite{IEEE02Zhu,IEEE06Uysal,IEEE06Djordjevic}. 
Our system utilizes an 810 nm wavelength for the quantum channel and a 660 nm wavelength for monitoring the beam deflections and correcting the quantum channel deflections.
To characterize the behavior of the beam wander caused by atmospheric turbulence depending on the wavelength, we measure the instantaneous x- and y-positions of two aligned laser beams at 810 nm and 660 nm, respectively, which were subjected to the same atmospheric conditions (Fig. \ref{fig:fig3}(a)). 
The two beams travel through the same optical path and enter the receiver telescope.
After reflcetion using an FSM (PI, S-330.4SL) with an angle resolution of 0.25 $\mu$rad and resonance frequency of 1.6 kHz, the beams split into PSD1 and PSD2. 
The inset in Fig. \ref{fig:fig3}(a) shows the active area of the pin-cushion-type PSD (Hamamatsu, C10443-03) \cite{PSD}.
The PSD sensor consists of a common cathode and four anodes, the optical signal on the PSD surface generating a photocurrent proportional to the light energy.
The photocurrent is split into the four anodes via a resistive layer, and the optical spot position $(x,y)$ on the PSD surface is extracted using the following expressions:
\begin{equation}
\begin{split}
x&=\frac{(V_2 + V_3)-(V_1+V_4)}{V_1+V_2+V_3+V_4}\times \frac{L_x}{2},\\
y&=\frac{(V_2 + V_4)-(V_1+V_3)}{V_1+V_2+V_3+V_4}\times \frac{L_y}{2},
\label{eq2}
\end{split}
\end{equation}
where $L_x$ and $L_y$ are the sensor dimensions (14$\times$14 mm$^2$), and $V_j$ is the voltage obtained from the anode current through the amplifiers. 
Figure \ref{fig:fig3}(b) presents the sample results for the beam-wander effects of the two beams over a period of 10 s, both signals following the same movements.

In addition to the wavelength dependence of the beam wander, we examined the beam-path drift owing to the outdoor temperature. 
As shown in Fig. \ref{fig:fig1}, because the active reflector is set up on an outdoor hill, the optical beam path along the free-space channel can be affected by the atmosphere's change in refractive index change and thermal expansions of the optical components due to changes in the outdoor temperature.
Figure \ref{fig:fig4} shows the influence of the outdoor temperature on the temporal beam drifts over a period of 400 min.
The measurements were performed on a bright sunny day from 11:00 to 17:40 on 27 April 2022 (Korea Standard Time, KST).
The temperature ranged from 12 $^{\circ}$C to 20 $^{\circ}$C.
For more clear observation of the outdoor temperature effect, a sunny day with a relatively big change in the temperature was chosen.
The long-term drift depends on the temperature variation.
During the night when the temperature doesn't change much, the long-term drift reduced significantly.
Therefore, as the sunlight becomes stronger, the temporal drift of the beam centroid position or radial distance increases, finally moving beyond the bounds of the telescope.
In this work, the long-term beam drift due to changes in the outdoor temperature is corrected by a coarse alignment using a wireless link between PSD1 and the hexapod supporting the target mirror in the outdoor active reflector, as shown in Fig. \ref{fig:fig5}(a).

\section{Results}
\subsection{Closed-loop feedback system}
The atmosphere in a free-space optical channel can lead to various detrimental effects$-$such as beam wander and slow temporal beam drift$-$which reduce the SNR and increase the QBER.
To compensate for atmospheric effects and to achieve stable SMF coupling for the quantum channel, we used coarse alignment followed by fine alignment using a closed-loop feedback system in which an FSM is placed before the quantum and tracking channels \cite{IEEE18Fernandez}.
This allows for the correction of the beam variations using the FSM, making it possible to monitor the beam correction in real-time.
The overall feedback system includes a control PC, FSM, PSD, and field programmable gate array (FPGA)-based  board for data processing as shown in Fig. \ref{fig:fig5}(a).
The 810 and 660 nm signals are received and divided using the DM and directed to the SMF and PSD1, respectively. 
Here, the distance of the SMF and PSD1 to the FSM must be equal to guarantee efficient beam correction for the quantum channel.
The FPGA board receives voltage information (V$_1$, V$_2$, V$_3$, and V$_4$) from PSD1, calculates the current beam position, and proviodes V$_x$ and V$_y$ output voltages (0$-$10 V), corresponding to the beam offsets from the center of PSD1 to the FSM. 
The control PC monitors the FPGA output signals and controlls the PID values at a sampling speed of 20 Hz using LabView.   
Coarse alignment (or slow feedback) is based on the use of a wireless bridge (TP-Link, CPE510) connecting PSD1 to the hexapod stage (PI, H-810.D2) in the outdoor active reflector.
A long-term beam drift due to the outdoor temperature is relatively slow compared to the beam wander, which can be as fast as 100$-$200 Hz \cite{MOTL16Casado}.
From the beam position value averaged over 10 s, a coarse alignment controlled the beam position to the center of the PSD1 every minute.
Immediately after the coarse alignment, a fine alignment (or fast feedback) is implemented at approximately 200 Hz using the FSM, driven by PID control through the FPGA.
Figure \ref{fig:fig5}(b) shows the beam centroid positions of the 660 nm tracking laser with a 250 m FSO link. 
The beam-wander range was reduced by the rapid feedback (or fine alignment). 
The inset presents the corresponding normalized histograms, and the Gaussian fit curves for the data are shown in Fig. \ref{fig:fig5}(c).
Consequently, the FWHM of the beam wander range was reduced from 350 to 24 $\mu$m.
To improve the fiber-coupling performance in the quantum channel, we can consider the chromatic aberration of the receiving telescope.
In the system, two laser beams with different wavelengths are received through a commercial Galilean telescope with $\times$20 magnification, which results in axial and lateral chromatic aberrations.
If we use a Cassegrain telescope as a receiver, it would eliminate these chromatic effects.

\subsection{250 m free-space optical link}
A quantum-correlation-based free-space optical link over 250 m was constructed based on a closed-loop feedback configuration, as shown in Fig. \ref{fig:fig5}.
To monitor the coupling efficiency into the SMF for the laser power immediately after the receiving telescope, a weak 810 nm laser ($\sim$ 10 mW) was used for measurements instead of the signal photons since it's impossible to measure the single photon counts directly in front of the SMF coupler.
We achieved an average coupling efficiency of 19 $\%$ as shown in Fig. \ref{fig:fig6}(a).
The inset shows the SMF coupling efficiency as the function of time.
Here, it shows the relatively large standard deviation ($\sim$ 520 $\mu$W) of the coupling efficiency.
This is due to the chromatic abberation effects. 
The large wavelength difference between the signal and the guiding laser and the use of the refractive receiving telescope cause axial and lateral chromatic abberations which lead to the misalignment of 810 nm and 660 nm wavelengths.
If we use a Cassegrain-type telescope as a receiver, these chromatic effects would be suppressed.
Figure \ref{fig:fig6}(b) shows the cross-correlation function between the idler and signal photons transmitted in a 250 m FSO channel.
The normalized cross-correlation function is defined as
\begin{equation}
g^{(2)}_{si}(\tau)= \frac{\langle E^{\dag}_s(t)E^{\dag}_i(t+\tau)E_i(t+\tau)E_s(t)\rangle}
{\langle E^{\dag}_s(t)E_s(t)\rangle\langle E^{\dag}_i(t+\tau)E_i(t+\tau)\rangle},
\label{eq3}
\end{equation}
where indices $s$ and $i$ represent the signal and idler photons, respectively.
$g^{(2)}_{si}(\tau)$ determines the rate of coincidence detection between modes $s$ and $i$ at a time delay $\tau$, and is useful for charaterizing light sources.

The blue $g^{(2)}_{si}(\tau)$ peak in Fig. \ref{fig:fig6}(b) represents the cross-correlation between the idler and signal photons immediately before the transmitted telescope.
The red data indicate the $g^{(2)}_{si}(\tau)$ function between the idler and received signal photons which were reflected from the outdoor active reflector.
The fiber-coupled signal photon counts was about 1$\times$10$^6$ counts/s immediately after the SPDC source, and the signal photon counts at the receiver site after the 250 m free-space propagation was about 5.5$\times$10$^4$ counts/s.
The background noise data are subtracted from the original $g^{(2)}_{si}(\tau)$ data.
The total system loss ($>$ 90 $\%$) includes the free-space channel loss ($\sim$16 $\%$), the SMF coupling loss ($\sim$81 $\%$) in the quantum channel, and the tranceiver optical stage loss ($\sim$45 $\%$).
The reduction in the $g^{(2)}_{si}(\tau)$ peak value is due to the total loss of the signal photons and daylight background noise (10 times larger than the number of signal photons).
For large daylight background noise, if we use a single-mode pump laser, linearly polarized signal photons and a black shielding box, the background noise would be further reduced using a narrower optical filter and filtering randomly polarized noise photons.

\section{Conclusions}
We have experimentally demonstrated a quantum correlation-based free-space optical link over 250 m, where the signal photons were sent to an outdoor target mirror, that was 125 m from the transceiver station. 
Based on a closed-loop feedback system with a 660 nm tracking laser, the atmospheric turbulence effects were well suppressed in the performance of the SMF coupling.
The temporal long-term drift due to the outdoor temperature was compensated by a coarse alignment (or slow feedback) that used the wirelessly-controlled hexapod stage of the target mirror.
The beam wander, which was faster than the temporal long-term drift, was also corrected by a fine alignment (or fast feedback) involving an FSM, driven by PID control, operating in response to the positional information of the PSD. 
After the receiving telescope, we achieved an average coupling efficiency into the SMF of around 19 $\%$.
We also measured the cross-correlation function $g^{(2)}_{si}(\tau)$ between the idler and signal photons transmitted through a 250 m free-space channel.
Although the transmittance photon loss and strong daylight degraded the peak value of $g^{(2)}_{si}(\tau)$, we could obtain sufficiently high peak values ($>$ 45) in our system owing to the strong temporal correlations between the photon-pairs.
Finally, the use of a Cassegrain telescope as a receiver and a more stringent filtering system will improve the performance of our quantum correlation-based FSO link system. 
Our work provides useful knowledge for designing an unmanned FSO link for urban quantum communication or retro-reflective quantum communication links which requires improved control of beam paths in outdoor conditions.

\end{document}